\documentclass[prl,twocolumn,twoside,showpacs,byrevtex,superscriptaddress]{revtex4}

\usepackage{graphicx,epsfig,verbatim,enumerate}
\usepackage{amssymb,amsmath}
\usepackage{ifthen}
\newboolean{twocolswitch}

\newcommand{\www}[1]{\url{#1}}
\newcommand{\req}[1]{(\ref{#1})}
\newcommand{\Req}[1]{Eq.~(\ref{#1})}

\newcommand{\dee}[1]{\mbox{d}#1}

\newcommand{\dstar}{d^\ast}
\newcommand{\dstari}{d_i^\ast}

\newcommand{\phifix}{{\phi^{\ast}}}

\newcommand{\phifixb}{{\phi_b^{\ast}}}

\newcommand{\dstardist}{g}

\setboolean{twocolswitch}{true}

\begin{document}

\title{
Universal behavior in a generalized model of contagion\\
{\small (To appear in Phys. Rev. Lett., 2004.)}
}

\author{
\firstname{Peter Sheridan}
\surname{Dodds}
}
\email{peter.dodds@columbia.edu}
\affiliation{Institute for Social and Economic Research and Policy,
  Columbia University,
  420 West 118th Street,
  New York, NY 10027.}

\author{
  \firstname{Duncan J.}
  \surname{Watts}
  }
\email{djw24@columbia.edu}
\affiliation{Department of Sociology,
  Columbia University,
  2960 Broadway,
  New York, NY 10027.}
\affiliation{
        Santa Fe Institute, 
        1399 Hyde Park Road, 
        Santa Fe, NM 87501.}

\markboth{Generalized Contagion}
{P.S.\ DODDS, D.J.\ WATTS}

\date{\today}

\begin{abstract}
Models of contagion arise broadly both in the biological and social
sciences, with applications ranging from the transmission of infectious
diseases to the diffusion of innovations and the spread of cultural fads.
In this Letter, we introduce a general model of contagion which,
by explicitly incorporating memory of past exposures to, for example,
an infectious agent, rumor, or new product, includes the main features of 
existing contagion models and interpolates between them.  
We obtain exact solutions for a simple version
of the model, finding that under general conditions only three classes
of collective dynamics exist, two of which correspond to 
familiar \textit{epidemic threshold} and \textit{critical mass}
dynamics, while the third is a distinct intermediate case.  
We find that for a given length of memory, the class into which a particular system
falls is determined by two parameters, each of which ought to be measurable empirically.  
Our model suggests novel measures for assessing the susceptibility of a
population to large contagion events, and also a possible strategy for
inhibiting or facilitating them.
\end{abstract}


\maketitle

Defined broadly as the transmission of an influence from
one individual to another, the concept of contagion occupies an
important place both in biology---specifically in mathematical
epidemiology~\cite{bailey1975,anderson1991a}---and in the
social sciences, where it is in manifested in problems as diverse as
the diffusion of innovations~\cite{bass1969,rogers1995}, the spread of
cultural fads~\cite{bikhchandani1992,banerjee1992,bikhchandani1998}, and the outbreak of 
political~\cite{lohmann1994} or social~\cite{granovetter1978} unrest.  

Despite the wide range of social and biological phenomena to which they have been applied,
existing models of contagion typically fall into one
of two categories that we delineate in terms of the relationship between
successive exposures of a ``susceptible'' to an ``infectious'' individual:
(1) what we call ``Poisson'' models, in which successive contacts result in
contagion with independent probability $p$; and (2) ``threshold'' models,
in which the probability of infection changes rapidly from low to
high as a critical number of simultaneous exposures is exceeded 
(thus the effect of any
single exposure depends strongly on the number of other exposures). 
The SIR model~\cite{kermack1927}, the canonical model of biological
contagion, is an example of a Poisson model, as is the
oft-cited Bass model~\cite{bass1969} from the diffusion of
innovations literature.  By contrast, numerous models in
sociology~\cite{granovetter1978}, economics~\cite{morris2000}, and political
science~\cite{schelling1973}, are explicitly threshold models; while others
still~\cite{glance1993,arthur1993,durlauf2001} embed thresholds
implicitly through the relative costs and benefits of one action versus
another.  

None of these models, however, treat the interdependencies
between exposures themselves as an object of study---rather they are simply
assumed to either exist or not exist---hence their effects on the
collective dynamics of contagion are unknown.  Furthermore, if, as we show
below, these effects turn out to be considerable, existing models provide
no way to determine the conditions under which one kind of collective
behavior or another should be expected.

\begin{figure*}[tbp]
  \centering
  \epsfig{file=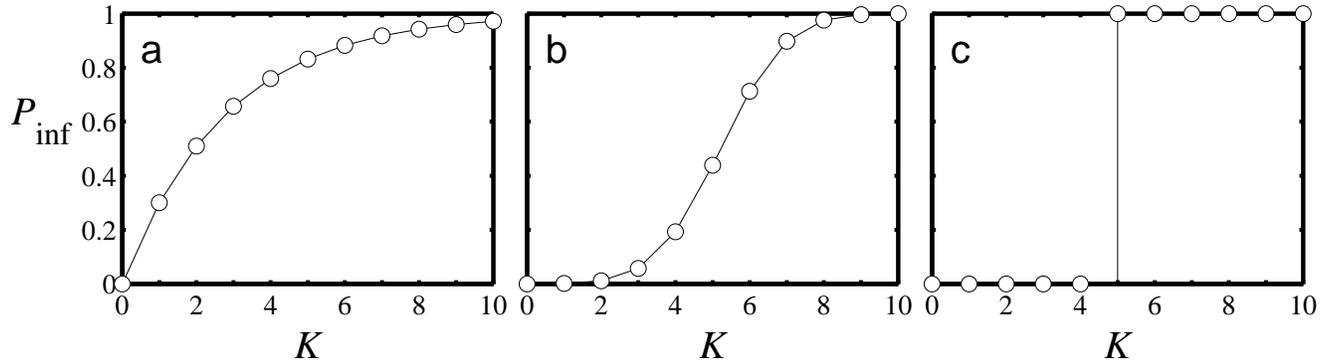,width=0.975\textwidth}
  \caption{
    Dose response curves for $T=10$ (see \Req{gcont-v2.eq:Pinf}):
    (a) Poisson (e.g., SIR-type)
    model: probability of receiving a positive dose by contacting an infective $p=0.3$, 
    distribution of dose sizes $f(d) = \delta(d-1)$,
    and distribution of individual dose thresholds $\dstardist(\dstar) = \delta(\dstar-1)$; 
    (b) Stochastic
    threshold model: $p=1$, $f(d)$ is distributed lognormally with unit mean
    and variance $0.4333$, and $\dstardist(\dstar) = \delta(\dstar-5)$;
    and
    (c) Deterministic threshold model: $p=1$, $f(d) =
    \delta(d-1)$, and $\dstardist(\dstar) = \delta(\dstar-5)$.
    }    
  \label{gcont-v2.fig:doseresponse}
\end{figure*}

In this Letter, we explore a generalized model of
contagion that, by introducing memory of past exposures to a contagious
influence, generalizes and interpolates between Poisson and threshold
models of contagion.  Our model is defined as follows.  Consider a population of $N$
individuals, each of which is in one of three states $S$ (susceptible), $I$
(infected), or $R$ (removed).  At each time step $t$, each individual $i$
comes into contact with one other individual $j$, drawn randomly from the
population.  If $i$ is susceptible and $j$ is infected then, with
probability $p$, $i$ receives a positive dose $d_i(t)$, drawn randomly from
some distribution of dose size $f(d)$; otherwise, $d_i(t)=0$.  Each
individual maintains a memory of doses received over the previous $T$ time
steps, recording a cumulative dose $D_i(t)=\sum_{t'=t-T+1}^{t} d_i(t')$. 
Susceptible individuals become infected if $D_i(t) \geq \dstari$ where
$\dstari$ ($i$'s \textit{dose threshold}) is drawn randomly at $t=0$ from a
distribution $\dstardist(\dstar)$, and remains fixed thereafter.  The
probability that a susceptible individual who encounters $K \le T$ infected
individuals in $T$ time steps will themselves become infected is therefore
\begin{equation}
  \label{gcont-v2.eq:Pinf}
  P_{\rm inf}(K)
  = 
  \sum_{k=1}^{K}
    \binom{K}{k}
    p^k
    (1-p)^{K-k}
    P_k,
\end{equation}
where
\begin{equation}
  \label{gcont-v2.eq:P_k}
  P_k
  = 
  \int_{0}^{\infty} \dee{\dstar}
  \dstardist(\dstar)
  P\left(
    \textstyle{\sum_{i=1}^{k} d_i \ge \dstar}
    \right)
\end{equation}
is the average fraction of individuals infected after receiving
$k$ positive doses in $T$ time steps, 
and $P(\sum_{i=1}^{k} d_i \geq \dstar)$ is the probability 
that the sum of $k$ doses drawn from $f(d)$ 
exceeds a given $\dstar$.

Equation~\req{gcont-v2.eq:Pinf} can be thought of as an average 
dose-response relationship~\cite{haas2002} for the
population in question.
Figure~\ref{gcont-v2.fig:doseresponse} displays three examples 
of~\Req{gcont-v2.eq:Pinf} for
different choices of $p$, $f(d)$, and $\dstardist(d_*)$.  
When all doses $d_i=\bar{d}$ are identical,
all members of the population have the same threshold $\dstar=\bar{d}$
and $p < 1$, then~\Req{gcont-v2.eq:Pinf} reduces to
the standard SIR model, Fig.~\ref{gcont-v2.fig:doseresponse}(a); 
and when $p=1$ and $\dstar>\bar{d}$, it is equivalent either to a 
stochastic~\cite{glance1993,durlauf2001},
or 
deterministic~\cite{granovetter1978,morris2000,watts2002}
threshold model, 
depending on whether doses are allowed to vary, Fig.~\ref{gcont-v2.fig:doseresponse}(b),
or are identical,
Fig.~\ref{gcont-v2.fig:doseresponse}(c).
More complicated choices of $f(d)$ and $\dstardist(\dstar)$ correspond to
many other kinds of models that
incorporate varying degrees
of interdependency between contagion events
and also heterogeneity across individuals.
 
Once infected, individuals may recover with probability $r$ if
$D_i(t)$ falls once more below $\dstar$ (otherwise they remain infected),
and recovered individuals become 
re-susceptible with probability $\rho$. While the resulting dynamics
is, in general, quite complex, in the 
special case of $\rho=1$ and $r=1$ (analogous to so-called 
SIS-type~\cite{pastor-satorras2001,kuperman2001}
models with instantaneous recovery), we can
write down an equation for the steady-state
fraction of infectives (fixed points) in the population~\footnote{The model possesses no periodic or chaotic solutions.}:
\begin{equation}
  \label{gcont-v2.eq:PhiStar}
  \phifix
  =
  \sum_{k=1}^{T}
  \binom{T}{k}
  (p\phifix)^{k}
  (1-p\phifix)^{T-k} 
  P_k,
\end{equation}
with $P_k$ as defined by~\Req{gcont-v2.eq:P_k}.

\begin{figure*}[tbp!]
  \begin{center}
    \epsfig{file=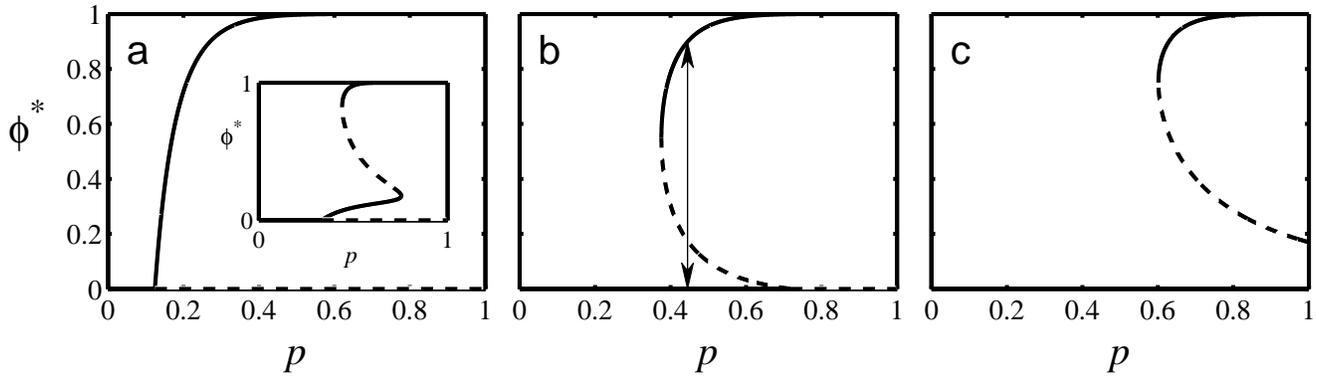,width=0.975\textwidth}
    \caption{
      Example fixed point curves for the three basic classes of contagion models
      (all curves are obtained numerically using the closed form expression
      of~\Req{gcont-v2.eq:PhiStar} and agree with results from simulations 
      (not shown)~\cite{dodds2004pb}):
      a) Class I: Epidemic threshold,
      b) Class II: Vanishing critical mass, and
      c) Class III: Pure critical mass.
      Dose sizes are lognormally distributed with mean 1 and variance $0.433$, $T=10$,
      and thresholds are uniformly set at
      a) $\dstar=0.5$,
      b) $\dstar=1.6$, and
      c) $\dstar=3$.
      In the class II example of (b), 
      trajectories of two initial conditions
      are indicated by arrows.  For $p=0.445$ and $\phi(0)=0.174$,
      the contagion fails to persist while 
      for the same $p$ and a slightly higher
      initial level of infection,
      $\phi(0)=0.175$, the contagion is
      sustained with 90\% of individuals eventually infected.
      At the transition between classes I and II, the saddle-node and
      transcritical bifurcations coincide yielding a continuous phase
      transition distinct in nature from those of class I.  The behavior of the fixed
      point curve near $p=p_c$ is $\phifix \propto (p-p_c)^{1/2}$ rather than
      $\phifix \propto (p-p_c)^{1}$,
      though in highly special cases the exponent is lower~\cite{dodds2004pb}.
      Inset in (a): An example of a more complicated fixed point diagram.
      Here, $T=20$, dose size is set to unity, $f(d)=\delta(d-1)$,
      and $\dstar=1$ with probability $0.15$ and $6$ with
      probability $0.85$.  
      In principle, systems with more bifurcations
      are possible although unlikely since a strongly 
      multimodal distribution of thresholds and/or dose sizes is required.
      }
    \label{gcont-v2.fig:classes}
  \end{center}
\end{figure*}

Exact solutions of \Req{gcont-v2.eq:PhiStar} may be obtained numerically 
(i.e., to arbitrary precision) and agree with results 
from our simulations~\cite{dodds2004pb}.
Moreover, we are able to determine analytically that the equilibrium behavior of our generalized
model falls into one of only three universal classes, 
examples of which are given in Fig.~\ref{gcont-v2.fig:classes}. 
We define these classes by the behavior of the fixed
point curves around a single transcritical bifurcation~\cite{strogatz1994},
which is present for all models and located at $p=p_c=1/(TP_1)$ and $\phifix=0$.
We find that for a given value of the memory parameter $T$,
the class into which a particular model falls depends only on the
values of two quantities: $P_1$ and $P_2$, the probability that
an individual will become infected as a result of one
and two exposures respectively.  
The three classes of behavior and their associated conditions are as follows.

\textit{Class I: epidemic threshold models}.  When $P_1 \geq P_2/2$, we find
the equilibria of Equation 2 fall on a bifurcation curve similar to that
shown in Fig.~\ref{gcont-v2.fig:classes}(a), in which 
the sole bifurcation is the transcritical one
at $p=p_c<1$ (analogous to a continuous phase transition~\cite{goldenfeld92}).
When $p\leq p_c$, a stable equilibrium
exists at $\phifix=0$ (i.e., all initial seeds die out exponentially fast),
and when $p>p_c$, $\phifix>0$ implying that a finite fraction of the
population will become infected (i.e., an epidemic will occur).  
Thus the dynamics of models in class I is qualitatively equivalent to that of
SIR-type models in which $p_c$, sometimes called the epidemic
threshold~\cite{murray2002}, determines the critical value of the
infectiousness $p$ required in order for an initial seed of infectives to
trigger an epidemic.  We therefore call models in class I
\textit{epidemic threshold} models.  

\textit{Class II: vanishing critical mass models}.  When $P_2/2 > P_1 \geq
1/T$, we find, as shown in Fig.~\ref{gcont-v2.fig:classes}(b), 
a change in the nature of the transcritical bifurcation.
Whereas for class I models, a stable fixed point curve emanates from $(p_c,0)$ with positive slope,
class II models have an unstable fixed point curve entering the transcritical bifurcation 
with a negative slope.
Accompanying this is the appearance of a saddle-node bifurcation~\cite{strogatz1994} at
$p=p_b<p_c$ and $\phi=\phifixb > 0$ (the system now exhibits a first-order phase transition).
The solid (dashed) lines in Fig.~\ref{gcont-v2.fig:classes}(b) correspond to the
position of stable (unstable) equilibria respectively; thus there
exists a region in $p$ in which two stable equilibria coexist, 
separated by an unstable equilibrium.
In other words, as shown by the arrows in Fig.~\ref{gcont-v2.fig:classes}(b), 
if the initial seed $\phi(0)$ falls below the unstable equilibrium,
then the contagion will die out; whereas if it falls above the unstable
equilibrium, then a large fraction of the population (corresponding to the
upper stable equilibrium) will be infected (i.e., the system is metastable~\cite{goldenfeld92}).
Thus infections in class II require a ``critical mass''~\cite{schelling1978} to succeed.  
Because the size of this critical mass decreases to zero for $p<1$, we call this class
of models \textit{vanishing critical mass} models, where we note that the
sensitivity to initial conditions implicit in critical mass dynamics is 
absent entirely from models in class I; hence the two classes are 
qualitatively distinct in terms of their equilibrium behavior.

\textit{Class III: pure critical mass models}.  When $1/T > P_1$, the
position of the transcritical bifurcation $p_c>1$; hence the only
bifurcation potentially remaining in the interval $0 \leq p \leq 1$ is the saddle-node
bifurcation, see Fig.~\ref{gcont-v2.fig:classes}(c).
Thus in the manner of first-order phase transitions, a finite critical
mass is always required (i.e., for all $p$) in order for an initial seed not to die out.
As a result, we call models in this class \textit{pure critical mass} models,
where our classification of pure and vanishing critical mass models
includes the dynamics of familiar threshold 
models~\cite{schelling1973,granovetter1978,watts2002},  
but is more general.

For some choices of $f(d)$ and $g(\dstar)$, additional equilibria are possible,
and correspond to additional saddle-node bifurcations in Figure 2.
Investigations of \Req{gcont-v2.eq:PhiStar} and numerical simulations~\cite{dodds2004pb}
indicate that for additional equilibria to appear, one or both of $f(d)$ and $g(\dstar)$ 
must be multimodal, with widely separated peaks. For example, the inset of
Fig.~\ref{gcont-v2.fig:classes}(a) shows the solutions of \Req{gcont-v2.eq:PhiStar},
where 15\% of the population have $\dstar=1$ and 85\% have $\dstar=6$.
In addition to the standard Class I transcritical bifurcation,
the inset shows a new saddle-node bifurcation.
Extra saddle-node bifurcations generated in this manner, do not affect either
the position or nature of the transcritical bifurcation, which
continues to depend only on $P_1$ and $P_2$.  
Hence our basic classification scheme, outlined above, 
is preserved for arbitrary $f(d)$ and $g(\dstar)$.

The existence of three basic classes, and also simple conditions
that determine which class a particular contagion model belongs to,
has interesting implications for understanding and
possibly influencing the collective dynamics of contagion, whether
biological or social.  First, the result
dramatically reduces the effective complexity of the generalized model by
showing that the qualitative dynamics are independent of
many of the details of the particular population [i.e., $f(d)$
and $\dstardist(\dstar)$]; hence our
use of the term ``universal classes''.  
Equally important, however, is that not all models fall into a single
class; that is, not all kinds of contagion are the same.  Furthermore, 
the three classes are non-degenerate in the rough sense that the conditions for 
each occupy broad regions of the parameter space.  This result is not to
suggest that some kinds of contagion are not more likely than others, but it 
does suggest that without empirical evidence about the relevant 
$P_1$ and $P_2$, little can be concluded about the 
corresponding collective dynamics.  

Because neither the mathematical epidemiology literature nor the
micro-organismal dose-response literatures have tended to question the
assumption that infection is a Poisson process, experimental evidence for
or against the assumption is limited.  In the context of biological
contagion, therefore, our model suggests a very clear test to validate or
refute the notion of interdependencies between exposures, and also
specifies the strength of interdependencies required in order for them to
be considered important with respect to the collective dynamics.  
Specifically,
if $P_1>P_2/2$ then the assumption of independence between successive
exposures can be considered valid, whether or not it is precisely true, in
the sense that the same qualitative behavior arises regardless.  If, however,
$P_1<P_2/2$ for some infectious agents or in some populations, then the 
interdependencies cannot be ignored, as their effect, in terms of the 
equilibrium states of the collective dynamics, may be dramatic.  

In the context of social contagion, our model has a different implication: that 
under very general, and one might argue likely, conditions, SIR-type models 
such as the Bass model that do not include memory or 
interdependencies between subsequent exposures, are incapable 
of capturing even the basic features of contagion dynamics.  However, 
there are also likely to be some applications in which the independence 
assumption does turn out to be valid---that is, threshold models are also unlikely 
to be universally appropriate models of social contagion---where the key 
point is that our model provides precise conditions under which one state 
or the other will pertain.  

A related point is that the conditions on $P_1$ and $P_2$ that separate the
three classes of behavior suggest novel intervention strategies for
suppressing, or alternatively stimulating, global contagion.  Assuming that
an individual's threshold $\dstari$ can be manipulated---for example,
increased (in a biological sense) by better preventative health treatment,
or decreased (in a social sense) by exerting some financial, social, or
cultural influence---then our results suggest that relatively minor manipulations 
(e.g., altering $P_1$ and $P_2$ to shift $p_c$ while leaving the class of contagion unchanged)
can have a dramatic impact on the ability of
a small initial seed to trigger a global contagion event (i.e., the accessibility
of a non-zero stable equilibrium) as well as on the size of such an event if it occurs.  
We note that manipulating $P_1$ and $P_2$
is not equivalent to manipulating $p$ since $p$ determines the relevant point
on the bifurcation curve, whereas $P_1$ and $P_2$ determine the shape of
the curve itself; one cannot subsume temporal interdependencies
within the independence assumption simply by lowering the per-event probability of
transmission.  Finally, we note that while recent theoretical work on  
controlling disease epidemics, or by contrast stimulating social 
contagion, has focused on individuals of exceptional influence
(e.g., so-called ``super-spreaders''~\cite{kemper1980,hyman2003}, 
and ``opinion leaders''~\cite{rogers1995}), our model suggests that it could 
be the most easily \textit{influenced} individuals (i.e., those contributing to $P_1$) who 
have the greatest impact on the dynamics of contagion.

\acknowledgments
The authors are grateful to Duncan Callaway for assistance with an
earlier version of this paper, and to
Paul Edelman, Charles Haas, Matthew Salganik, and Joshua Zivin for
helpful comments.  This research was supported in part by the National
Science Foundation (SES 0094162), the Office of Naval Research, Legg
Mason Funds, the Intel Corporation, and the James S.\ McDonnell Foundation.

\end{document}